# A relativistic third-order algebraic diagrammatic construction theory for electron detachment, attachment and excitation problems

Sudipta Chakraborty[1], Tamoghna Mukhopadhyay[1], Malaya K. Nayak[2,3†] and Achintya Kumar Dutta[1]*

1. *Department of Chemistry, Indian Institute of Technology Bombay, Powai, Mumbai 400076, India.*
2. *Theoretical Chemistry Section, Bhabha Atomic Research Centre, Trombay, Mumbai 400085, India.*
3. *Homi Bhabha National Institute, BARC Training School Complex, Anushakti Nagar, Mumbai 400094, India.*

**Abstract**

We present the theory and implementation of a highly efficient relativistic third-order algebraic diagrammatic construction [ADC(3)] method based on a four-component (4c) Dirac-Coulomb (DC) Hamiltonian for the calculation of ionization potentials (IP), electron affinities (EA), and excitation energies (EE). Benchmarking calculations for IP, EA, and EE were performed on both atomic and molecular systems to assess the accuracy of the newly developed four-component relativistic ADC(3) method. The results show good agreement with the available experimental data. The Hermitian nature of the 4c-ADC(3) Hamiltonian, combined with the perturbative truncation of the wave function, offers significant computational advantages over the standard equation-of-motion coupled-cluster approach, particularly for property calculations. The method's suitability for property calculations is further demonstrated by computing oscillator strengths and excited-state dipole moments for heavy elements.

† mknayak@barc.gov.in; mk.nayak72@gmail.com; mknayak@hbni.ac.in

*achintya@chem.iitb.ac.in

## 1. Introduction

The excited state energies and transition properties have vast applications in different branches of science. Simulation of these properties plays a significant role[1] in the field of spectroscopy, photochemistry, and photobiology. One of the fundamental challenges for excited state calculations is the development of theoretical methods that are accurate for all kinds of excited states and have low computational costs at the same time. Within the framework of a wavefunction-based method, one can simulate excitation energy in two different ways – I. " $\Delta$ -based methods" or the separate calculation of ground and excited states and taking the difference afterward or II. "direct-difference of energy-based methods" or the methods involving single calculation on an initial ground state to generate the excitation energy as an eigenvalue of a secular Hamiltonian. While the former suffers problems like symmetry breaking, variational collapse, etc., the latter is generally free from these problems and is more widely used. Moreover, the direct energy difference-based methods have an additional advantage over the former, providing access to the transition probability for each of the excited states. Among the various direct-difference based wave-function methods, the equation-of-motion coupled cluster (EOM-CC)[2–5] theory has emerged as one of the most popular options for the calculation of excitation energy. The linear response coupled cluster (LR-CC)[6,7] method gives identical results as the EOM-CC method for excitation energy, although both of them are derived from different theoretical point of views. However, due to the non-Hermitian nature of the coupled cluster similarity transformed Hamiltonian, the calculation of transition properties in both LR-CC and EOM-CC methods require significant additional effort over the energy calculation.

The Algebraic Diagrammatic Construction (ADC)[8–14] theory, being Hermitian, has a great advantage in terms of property calculation compared to non-Hermitian coupled cluster based excited state methods. In the non-Hermitian formalism, both left and right eigenvectors need to be obtained to calculate the excited state property.[5] On the contrary, in the Hermitian formalism like ADC, one needs to calculate only a single set of eigenvectors for each excited state for property calculations, which reduces the computational cost nearly to half. The ADC theory was first derived for the polarization propagator using diagrammatic perturbation expansion[11]. Later on, Intermediate State Representation (ISR)[9,10] became more popular due to its straightforward formalism, and easy access to the excited state wavefunction. The perturbative truncation in ADC based methods reduces its computational cost compared to the corresponding EOM-CC method. Mukherjee and Kutzelnigg[15] have shown an alternative derivation of ADC using effective Liouvillian formalism, which has been used in recent works of Sokolov and coworkers.[16,17]

For the simulation of excited states of heavy elements, one needs to include the relativistic effect in the calculation. One of the most accurate ways to incorporate the relativistic effect is to use a four-component Dirac-Coulomb (DC) Hamiltonian. The Dirac equation for the many electronic systems is generally solved using the Dirac-Hatree-Fock (DHF) approximation.[18] In recent times, the ADC methods have been extensively used for the energy and property[16,19–22] calculations of atoms containing lighter elements. Extension to the fourth-order ADC scheme (ADC(4)) has also been achieved in the literature[23–25]. In the relativistic domain, second-order and extended second-order versions of ADC have been implemented by Pernpointner et al[26,27] for excitation energies and transition dipole moments. The second-order ADC scheme often doesn't provide adequate accuracy, and one needs to go at least to third order in perturbation to get the desired accuracy in the calculations[28]. In this manuscript, we present the theory, implementation, and benchmark of the ionization potential, electron affinity, excitation energy, and excited state first-order properties within the four-component relativistic ADC(3) formalism.

## 2. Theory

### 2.1 Relativistic Algebraic Diagrammatic Construction (ADC) Theory

Although the algebraic diagrammatic construction was initially derived in the context of propagator theory,[11,14,24] we follow the intermediate state representation (ISR)[8–10] derivation of ADC. In order to form the intermediate state basis, firstly, one needs to operate a linear excitation/ionization/electron-attachment operator on the correlated ground state wave function to generate correlated excited states (CES)

$$|\psi_I^0\rangle = \hat{C}_I\,|\psi_0^N\rangle \tag{1}$$

The nature of the operator $\hat{C}_I$ depends upon the desired target state

For ionization:

$$\{\hat{C}_I\} = \{\hat{c}_i, \hat{c}_a^\dagger\hat{c}_i\hat{c}_j, \hat{c}_b^\dagger\hat{c}_a^\dagger\hat{c}_i\hat{c}_j\hat{c}_k, ...; i<j<k..., a<b<...\} \tag{2}$$

For excitation:

$$\{\hat{C}_I\} = \{\hat{c}_a^\dagger\hat{c}_i, \hat{c}_b^\dagger\hat{c}_a^\dagger\hat{c}_i\hat{c}_j, \hat{c}_c^\dagger\hat{c}_b^\dagger\hat{c}_a^\dagger\hat{c}_i\hat{c}_j\hat{c}_k, ...; i<j<k..., a<b<c...\} \tag{3}$$

For electron-attachment:

$$\{\hat{C}_I\} = \{\hat{c}_a^\dagger, \hat{c}_b^\dagger\hat{c}_a^\dagger\hat{c}_i, \hat{c}_c^\dagger\hat{c}_b^\dagger\hat{c}_a^\dagger\hat{c}_i\hat{c}_j, ...; i<j..., a<b<c...\} \tag{4}$$

These correlated excited states are generally not orthonormal, but they can be orthogonalized using Gram-Schmidt orthogonalization[9] to form precursor states. These precursor states are then allowed to undergo a symmetric orthonormalization to form the intermediate state basis. In the ISR representation, the ADC excited state wavefunction for the $K^{th}$ state can be represented in terms of the intermediate state basis as follows

$$\left|\Psi_K^{ex}\right\rangle = \sum_I \mathbf{Y}_{IK}\left|\tilde{\Psi}_I\right\rangle \tag{5}$$

The coefficient matrix $\mathbf{Y}_{IK}$ as eigenvectors and $\Omega$ as eigenvalues are obtained in the diagonalization of the secular matrix ($\mathbf{M}_{IJ}$) of the ADC Hamiltonian

$$\mathbf{MY} = \mathbf{Y}\Omega \tag{6}$$

where

$$\mathbf{M}_{IJ} = \langle\tilde{\Psi}_I\,|\,\hat{H} - E_0^N\,|\,\tilde{\Psi}_J\rangle \tag{7}$$

and

$$\mathbf{Y}_{IK} = \left\langle\tilde{\Psi}_I\middle|\Psi_K\right\rangle; \mathbf{Y}^\dagger\mathbf{Y} = \mathbf{1} \tag{8}$$

In ADC, this secular matrix ($\mathbf{M}_{IJ}$) is expanded in the perturbation order as

$$\mathbf{M} = \mathbf{M}^{(0)} + \mathbf{M}^{(1)} + \mathbf{M}^{(2)} + \mathbf{M}^{(3)} + ... \tag{9}$$

This series is truncated at perturbation order $n$ for ADC($n$) theory. For example, taking n =2 leads to the ADC(2) method and n=3 leads to the ADC(3) method and so on.

The four-component relativistic ADC method is based on the Dirac-Hartree-Fock (DHF) mean-field wave function calculated using the Dirac-Coulomb Hamiltonian. For the molecular systems, the Dirac-Coulomb Hamiltonian ($H_{DC}$) can be written as,

$$H_{DC} = \sum_{i}^{N}\left[ c\vec{\alpha}_i.\vec{p}_i + \beta_i m_0 c^2 + \sum_{A}^{N_{nuc}} V_{iA} \right] + \sum_{i<j}^{N}\frac{1}{r_{ij}} I_4 \tag{10}$$

Here, $m_0$ is the rest mass of the electron and $c$ is the speed of light. $\vec{p}_i$ and $V_{iA}$ are the momentum and potential energy operators, respectively, for the $i^{th}$ electron in the field of the nucleus $A$. The $\alpha$ and $\beta$ are Dirac matrices and $I_4$ is a $4\times4$ identity matrix.

For many-body systems, the Dirac-Hartree-Fock (DHF) equation can be written in the matrix form as

$$\begin{bmatrix} \hat{V}+\hat{J}-\hat{K} & c(\sigma_{psm}.\hat{p})-\hat{K} \\ c(\sigma_{psm}.\hat{p})-\hat{K} & \hat{V}-2m_0c^2+\hat{J}-\hat{K} \end{bmatrix}\begin{bmatrix} \psi^L \\ \psi^S \end{bmatrix} = E\begin{bmatrix} \psi^L \\ \psi^S \end{bmatrix} \tag{11}$$

Here, $\left|\psi^L\right\rangle$ and $\left|\psi^S\right\rangle$ denote the large and the small components of the four component spinor $\left|\psi\right\rangle$ respectively, $\hat{V}$ denotes the electron-nuclear interaction, $\hat{J}$ and $\hat{K}$ represent the Coulomb and the exchange operators respectively and $\sigma_{psm}$ denotes Pauli spin matrices. After the solution of the DHF equation, the integrals in a molecular spinor basis are generated using the no-pair approximation.[18] The ADC equations are generally solved using the Davidson iterative diagonalization procedure[29], where the trial vectors are multiplied with the dressed Hamiltonian to give the so-called "sigma" vectors. The programmable expressions for the ADC(2) and ADC(3) methods for the ionization energy, electron attachment and excitation energy have been provided in the Supplementary Material. In the present implementation, Kramers restriction[18] has not been used to keep it independent of the framework used for the generation of the underlying DHF spinors.

## 2.2 Excited state property and transition dipole moment

The ISR formalism discussed in the previous section can also be extended to general operators besides the Hamiltonian. This gives the ISR-ADC method complete versatility, enabling applications that were not feasible with original polarisation propagator formalism[30]. Consider a one-particle Hermitian operator $\hat{D}$, associated with a physical property of interest and can be represented in the second quantization form as

$$\hat{D} = \sum_{pq} d_{pq}\{c_p^{\dagger}c_q\} \tag{12}$$

where $d_{pq}$ are the matrix elements of $\hat{D}$, constructed in the Dirac-Hartree-Fock (DHF) reference basis $\{\chi_p\}$,

$$d_{pq} = \left\langle \chi_p \middle| \hat{D} \middle| \chi_q \right\rangle \tag{13}$$

The calculation of an excited-state first-order property associated with a particular operator $\hat{D}$ involves evaluating the respective expectation value defined as

$$D_K = \left\langle \Psi_K^{ex} \middle| \hat{D} \middle| \Psi_K^{ex} \right\rangle \tag{14}$$

The orthogonal ground state reference wavefunction is not considered in the ISR, so the expectation value of the ground-state shifted operator $(\hat{D} - \hat{D}_0)$ in the ISR formalism can be determined by,

$$\begin{aligned} \bar{D}_K &= \sum_{IJ} \mathbf{Y}_{IK}^* \langle \tilde{\psi}_I \,|\, \hat{D} - \hat{D}_0 \,|\, \tilde{\psi}_J \rangle \mathbf{Y}_{JK} \\ &= \mathbf{Y}_K^\dagger \mathbf{G} \mathbf{Y}_K \end{aligned} \tag{15}$$

where, $\mathbf{G}$ is the ISR of $\hat{D}$ and associated matrix elements can be expressed as

$$G_{IJ} = \langle \tilde{\Psi}_I \,|\, \hat{D} \,|\, \tilde{\Psi}_J \rangle \tag{16}$$

Finally, in order to get back to the total expectation value of the K$^{th}$ excited state, one needs to add the ground state component to the transition contribution provided by $\bar{D}_K$ in Eq. (15), and so, the expression of the total expectation value of the K$^{th}$ excited state is modified to,

$$\begin{aligned} \bar{D}_K &= \sum_{IJ} \mathbf{Y}_{IK}^* \langle \tilde{\psi}_I \,|\, \hat{D} - \hat{D}_0 \,|\, \tilde{\psi}_J \rangle \mathbf{Y}_{JK} + \sum_{IJ} \mathbf{Y}_{IK}^* \delta_{IJ} \hat{D}_0 \mathbf{Y}_{JK} \\ &= \mathbf{Y}_K^\dagger \mathbf{G} \mathbf{Y}_K + \hat{D}_0 \end{aligned} \tag{17}$$

Besides the excited state first-order property, another important property is the prediction of oscillator strength or spectral intensities, for which one needs to consider the transition dipole moment between two states. The ground-to-excited state transition moment can be defined as,

$$T_{0\to K} = \langle \Psi_K \,|\, \hat{D} \,|\, \Psi_0 \rangle \tag{18}$$

Similarly, in the ISR formalism, it can be expressed as,

$$T_{0\to K} = \sum_K \mathbf{Y}_{IK}^* F_I = \mathbf{Y}_K^\dagger \mathbf{F} \tag{19}$$

where,

$$F_I = \sum_I \mathbf{Y}_{IK}^* \langle \tilde{\Psi}_I \,|\, \hat{D} - \hat{D}_0 \,|\, \Psi_0 \rangle \tag{20}$$

The **F** and **G** can be expanded in terms of the second quantized operators as,

$$F_I = \sum_{pq} f_{I,pq} d_{pq} \quad \text{and} \quad \mathrm{G}_I = \sum_{pq} g_{I,pq} d_{pq} \tag{21}$$

where,

$$f_{I,pq} = \langle \tilde{\Psi}_I \,|\, c_p^\dagger c_q \,|\, \Psi_0 \rangle \quad \text{and} \quad g_{I,pq} = \langle \tilde{\Psi}_I \,|\, c_p^\dagger c_q \,|\, \tilde{\Psi}_J \rangle \tag{22}$$

One can describe the one-particle excited state reduced density matrix $(\rho_{pq}^K)$ as

$$\rho_{pq}^K = \left\langle \Psi_K^{ex} \middle| c_p^\dagger c_q \middle| \Psi_K^{ex} \right\rangle = \sum_{IJ} \mathbf{Y}_{IK}^* \langle \tilde{\Psi}_I \,|\, c_p^\dagger c_q \,|\, \tilde{\Psi}_J \rangle \mathbf{Y}_{JK} = \mathbf{Y}_K^\dagger \mathbf{g}_{pq} \mathbf{Y}_K \tag{23}$$

and ground-to-excited state transition density matrix $(\rho_{pq}^{0\to K})$

$$\rho_{pq}^{0\to K} = \left\langle \Psi_K^{ex} \middle| c_p^\dagger c_q \middle| \Psi_0 \right\rangle = \sum_I \mathbf{Y}_{IK}^* \langle \tilde{\Psi}_I \mid c_p^\dagger c_q \mid \Psi_0 \rangle = \mathbf{Y}_K^\dagger \mathbf{f}_{pq} \tag{24}$$

Contracting these one-particle reduced density matrices with the one particle property operator, one can evaluate the excited state and transition property according to

$$D_n = \left\langle \Psi_K^{ex} \middle| \hat{D} \middle| \Psi_K^{ex} \right\rangle = \sum_{pq} D_{pq} \rho_{pq}^K \tag{25}$$

$$T_{0\to K} = \left\langle \Psi_K^{ex} \middle| \hat{D} \middle| \Psi_0 \right\rangle = \sum_{pq} D_{pq} \rho_{pq}^{0\to K} \tag{26}$$

In this work, we have calculated the excited state dipole moment and transition dipole moment. In the case of ADC, the computational advantage is that because of its Hermitian nature, only one eigenvector per state needs to be calculated. One can obtain the square of the transition dipole moment as

$$\left|T_{0\to K}\right|^2 = \left|T_{0\to K}^{xx}\right|^2 + \left|T_{0\to K}^{yy}\right|^2 + \left|T_{0\to K}^{zz}\right|^2 \tag{27}$$

Subsequently, the oscillator strength (observable) corresponding to the ground to each excited state is as follows,

$$f_{osc}^{0\to K} = \frac{2}{3} \Omega^{0\to K} \left|T_{0\to K}\right|^2 \tag{28}$$

The G and F in equations (16) and (20) can also be expanded in perturbation order.

$$\mathbf{G} = \mathbf{G}^{(0)} + \mathbf{G}^{(1)} + \mathbf{G}^{(2)} + \ldots\ldots. \tag{29}$$

$$\mathbf{F} = \mathbf{F}^{(0)} + \mathbf{F}^{(1)} + \mathbf{F}^{(2)} + \ldots\ldots. \tag{30}$$

Figure 1 illustrates the perturbation order required for each block at different levels of ADC. In strict ADC(2) and ADC(3) variants, both the ADC vectors and the operator are expanded up to the second and third orders in perturbation theory, respectively. However, the excited state property and transition properties for the present implementation of the relativistic ADC(3) method have been calculated using a mixed ADC(3/2) scheme, where the third-order ADC(3) vectors are combined with a second-order ISR representation. Dreuw and co-workers[31] have recently reported an intermediate state representation, which is complete up to third order in perturbation in the context of non-relativistic ADC method. However, the strict third-order intermediate state representation has a higher computational cost and generally shows very little improvement over the ADC(3/2) method for property calculations[31].

### 3. Computational Details

Our in-house quantum chemistry software program BAGH[32], is used for all of the calculations. BAGH currently has three interfaces, namely PySCF[33–35], GAMESS-US[36] and DIRAC[37]. All the four-component relativistic ADC (4c-ADC) calculations presented in this paper have been calculated using the PySCF interface. The majority of BAGH is written in Python, with Cython and FORTRAN used to optimize the computationally demanding parts. The specific information about the used basis set and truncation of correlation space at the canonical level is discussed in detail in the results and discussion section.

## 4. Results and Discussion

### 4.1 Ionization potential

The ionization potential (IP) of halogen monoxide anions has grabbed special attention, especially in the field of atmospheric chemistry[38,39], for the last few decades. As one goes down the group in the periodic table, the relativistic effect becomes significant, and the relativistic ADC(3) method can be an efficient option for calculating the ionization potentials of these molecules. Table 1 presents the ionization potential (IP) values for the first two ionized states of $XO^-$ (where X = Cl, Br, and I), along with their splitting as calculated using IP-ADC(2) and IP-ADC(3) methods. One significant consequence of relativistic effects is the splitting of ionization spectra due to spin-orbit coupling. To accurately account for this spectroscopic splitting, it is essential to move beyond scalar relativistic effects and include spin-orbit coupling in the calculations. The 4c-ADC(3) method provides an efficient approach to incorporate both relativistic and electron correlation effects. The splitting of states for halogen monoxide anions ($XO^-$; X = Cl, Br, and I) was calculated using the Dyall.av3z basis set for the halogen atom (X) and an uncontracted aug-cc-pVTZ basis set for the oxygen atom. In the IP calculations, 26 occupied and 206 virtual spinors were considered in the correlation space for $ClO^-$, while 44 occupied and 246 virtual spinors were used for $BrO^-$. For $IO^-$, 62 occupied and 248 virtual spinors were included in the correlation calculations. All IP results for halogen monoxide anions were obtained at the experimental geometries of the corresponding halogen monoxides.

The splitting between $^2\Pi_{1/2}$ and $^2\Pi_{3/2}$ states increases from $ClO^-$ to $IO^-$ as the extent of spin-orbit coupling increases down the group. For all three systems, the calculated splitting were underestimated using the 4c-ADC(2) method, while they were overestimated using 4c-ADC(3) method. The magnitudes of errors in both cases are comparable, and the maximum deviation in splitting is observed for IO- with the magnitude of error being 0.13 eV, which is within the error bar of the ADC(3) method. This tendency of ADC(2) and ADC(3) to produce errors of similar magnitude but opposite sign has been previously documented by Dreuw and co-workers in the context of excitation energies.[40] As a potential solution they have suggested to scale the third order term with a single tuning parameters $x$ = 0.4–0.5. We call this method as 4c-IP-ADC[(2)+x(3)] and used a scaling parameter of x=0.5. One can see that the error in the splitting obtained for all the molecules at 4c-IP-ADC[(2)+x(3)] method is small and comparable to the 4c-IP-EOM-CCSD method.

### 4.2 Electron affinities

Unlike ionization potential, the electron affinity estimated at Koopman's level shows large errors, as the error due to missing orbital relaxation and correlation effect reinforces each other. Moreover, the effect of electron correlation plays a major role in determining the stability of the anions, and many of the anions are only bound when electron correlation effect is included in the calculations. In addition, the study of electron attachment in systems containing heavy elements requires the consideration of relativistic effects in the calculation.

We have used the four-component relativistic EA-ADC(3) method to investigate the vertical electron attachment energies of halogen atoms and compared them with the experimental values. The Table 2 presents electron affinity (EA) values for a series of halogen atoms (F, Cl, Br, I and At) calculated using four-component relativistic, spin-free exact two-component relativistic in its one electron variant (SFX2C1e) and non-relativistic (NR) Hamiltonian, alongside experimental values. For all the halogen atoms considered here the d-aug-Dyall.v4z basis set has been used. For F and Cl atoms, all the virtual spinors have been considered in the correlation space. Whereas, for Br, I and At, virtual spinors with energies only within 6344 a.u., 5000 a.u. and 816 a.u. are considered respectively for electronic correlation, which leads to 360, 380 and 430 virtual spinors for Br, I and At, respectively. Non-relativistic calculations result in EA values that show deviation from experimental results for lighter atoms like F and Cl, and the error increases significantly as we go down the group. All the 4c-ADC(3) variant significantly understates the EA of the F atom and as expected, the non-relativistic result shows a large error of 0.515 eV. The use of SFX2C1e Hamiltonian does not lead to any appreciable improvement over non-relativistic results. The 4c-ADC(2) method shows a relatively better agreement with the experiment results. The errors are even smaller in 4c-EA-EOM-CCSD method, but it still shows a deviation of 0.174 eV with respect to the experiment. For Cl as well, the EA value in NR-EA-ADC(3) is underestimated by 0.166 eV compared to the experimental value, whereas 4c-EA-ADC(3) calculations result in a slight overestimation of only 0.061 eV, indicating improved accuracy. The 4c-EA-EOM-CCSD method shows a slightly larger error of 0.12 eV. The use of SFX2C1e Hamiltonian gives a very similar result to that of the non-relativistic one for Cl atom also. For the Br atom, non-relativistic ADC(3) calculation overestimate the EA values by 0.028 eV, but the error gets reduced to 0.017 eV in the 4c-ADC(3) method. The 4c-EA-EOM-CCSD method shows a larger error of 0.142 eV. The SFX2C1e-ADC(3) method gives a slightly better agreement with experimental results compared to the non-relativistic one, although the values are very similar to each other. The first significant difference in electron affinity (EA) values between the SFX2C1e and NR-ADC(3) methods is observed for iodine (I), and this difference is even more pronounced for At atom. The error in EA-ADC(3) values gets reduced by 0.075 eV and 0.182 eV for I and At, respectively, on going from non-relativistic to SFX2C1e Hamiltonian. However, the error in the SFX2C1e-ADC(3) results compared to experimental values for I and At are 0.155 eV and 0.504 eV, respectively. The 4c-ADC(3) method shows an smaller error of 0.057 eV and 0.054 eV, respectively. This demonstrates that the inclusion of the scalar relativistic effect is not sufficient for the accurate prediction of electron affinity values of heavy elements, and one needs to include the spin-orbit coupling effects in the calculation. The 4c-EA-EOM-CCSD method shows an error of 0.13 eV and 0.133 eV with respect to the experiment for I and At.

## 4.3 Excitation Energies

Accurate simulation of excitation energies of atoms and molecules containing heavy elements requires balanced description of relativistic and electron correlation effects. Our implementation of EE-ADC(3) method in the full four-component relativistic framework enables us to perform a balanced inclusion of both electron correlation and relativistic effect. The triiodide anion ($I_3^-$) is a popular test to benchmark relativistic electron correlation methods for excited states. Several investigations have been performed for electronically excited states of triiodide ion in both gas[41–45] and solution phases[46–49], focusing on the complicated dissociation dynamics after excitation. The photofragment yield study by Neumark and coworkers[45] in the gas phase exhibits two bands near 360 nm (3.4 eV) and 290 nm (4.3 eV). Previously semiempirical theoretical studies by Okada et al.[50] also showed that these two bands are a combination of ${}^3\Pi_{0+u}$ and ${}^1\Sigma_{0+u}$ states. Several theoretical studies have been reported in the literature on triiodide[51] anion, including Intermediate Hamiltonian Fock-Space Coupled Cluster (IHFS-CC)[52,53], EOM-CCSD[54], CASPT2[55,56], MRCI[57–59], and TD-DFT[60] methods, in the relativistic framework. Wang and co-workers[61] have reported EOM-CCSD results for the same molecule with a perturbative inclusion of the SOC effect. We have compared our 4c-ADC(3) results for $I_3^-$excited states with previous relativistic calculations such as EE-EOM-CCSD, EE-IHFS, and CASPT2 methods based on four-component Dirac-Coulomb Hamiltonian, as well as SOC-EOM-CCSD results by Wang and co-workers[61]. Results obtained from the 4c-EE-ADC(3) are presented in Table 3, along with those obtained from the aforementioned methods. Following the work of Gomes and co-workers[54], a bond length of 2.93 Å and the Dyall.av3z basis set has been used for the calculation. All the spinors with energies ranging from -3.0 to 12.0 $E_h$ have been correlated, which involves 52 electrons and 332 virtual spinors. The IHFS-CCSD[53], being the most accurate among the methods compared, is used as the reference for statistical analysis, and the statistical parameters are presented in Table 3. On going from ADC(2) to ADC(3), the excitation energies for all the states undergoes a redshift and improves the agreement with IHFS-CCSD results. It can also be seen that excitation energies obtained from 4c-ADC(3) are closer to the IHFS-CCSD values than the 4c-EOM-CCSD method and show a smaller mean absolute deviation (MAD) than the 4c-EOM-CCSD method. However, the STD of both 4c-ADC(3) and 4c-EOM-CCSD method are comparable. The performance of the EOM-CCSD method with perturbative inclusion of SOC depends upon the level at which level the SOC interactions are included[61], with the variant (c) where the SOC operator is included only in the EOM level and the unliked terms were dropped from the sigma vectors gives best agreement with IHFS-CCSD results. It can be seen that the 4c-ADC(3) method gives the best agreement with the IHFS-CCSD results, among all the methods compared here.

In addition to the excitation energies, it is essential to correctly simulate the energy differences. The accurate determination of excitation energies and fine-structure splitting in ions is fundamental to understanding their electronic structure and spectral properties. For heavy elements like $Ga^+$, $In^+$, and $Tl^+$, relativistic effects become significant and must be considered in the calculation. Here we have employed the 4c-ADC(3) method to compute the excitation energies and fine-structure splitting of $Ga^+$, $In^+$, and $Tl^+$ ions. Triply augmented Dyall's valance triple zeta (t-aug-Dyall.v3z) basis set has been used for the calculation, and virtual spinors up to 2000 $E_h$ have been considered in the correlation space, along with the frozen core approximation for all the atoms. This leads to 278, 300 and 402 virtual spinors for $Ga^+$, $In^+$ and $Tl^+$, respectively. The calculated excited states corresponding to the three spin-forbidden transitions ${}^1S_0 \to {}^3P_0$ , ${}^1S_0 \to {}^3P_1$ and ${}^1S_0 \to {}^3P_2$ and one spin-allowed ${}^1S_0 \to {}^1P_1$ transition. The fine-structure splitting for these three excited states are shown in Table 4. For $Ga^+$, the computed excitation energies for all the four transitions are consistently underestimated compared to the experimental values and the deviation is lowest for the ${}^1S_0 \to {}^1P_1$ transition. In the case of $In^+$ and $Tl^+$ also, a similar underestimation compared to the experiment is observed. However, the fine-structure splittings are well reproduced. For the ${}^3P_0 \to {}^3P_1$ and ${}^3P_1 \to {}^3P_2$ states in $Ga^+$, the computed splitting is very close to the experimental value. For $In^+$ and $Tl^+$, the computed splittings are very close to the experimental results for ${}^3P_0 \to {}^3P_1$, and a slightly elevated error is observed for ${}^3P_1 \to {}^3P_2$ splitting ( 4.56% and 4.32% for $In^+$ and $Tl^+$, respectively). The magnitude of splitting increases while going down the group, and this pattern is clearly observed in the calculated splitting of $Ga^+$, $In^+$, and $Tl^+$. An increase of 0.077 eV and 0.307 eV in splitting for the ${}^3P_0 \to {}^3P_1$ is observed on going from $Ga^+$ to $In^+$ and $Tl^+$, respectively. Similarly, an increase of 0.180 eV and 0.814 eV is observed for the ${}^3P_1 \to {}^3P_2$ splitting. The trend is consistent with the available experimental values.

### 4.4 Excited state properties

The lowest optically active transition in closed-shell alkali metal cations (such as $Na^+$, $K^+$, $Mg^{2+}$ etc.) within the non-relativistic framework is typically due to the ${}^1S \to {}^1P$ transition. Consequently, in the non-relativistic picture, the absorption spectrum corresponding to the low-lying transition in alkali metal cations will show a single peak due to the transition ${}^1S_0 \to {}^1P_0$. In the relativistic framework, there is an additional low-intensity peak due to the transition ${}^1S_0 \to {}^3P_0$, which borrows intensity through the spin-orbit coupling. Figure 2 presents the absorption spectra corresponding to the low-lying excited state of the $K^+$ atom in the relativistic and non-relativistic ADC(3) methods. The s-aug-Dyall.v2z basis set has been used for the calculations. One can see that the 4c-ADC(3) method indeed shows two peaks, whereas the low-intensity peak is missing in the non-relativistic variant.

To further investigate the performance of 4c-ADC(3) method, we have calculated the transition dipole moment corresponding to the first four bright states of Xe. The d-aug-Dyall.ae3z basis set has been used for the calculations. The frozen core approximation and a virtual spinor cut-off of 361 $E_h$ have been used which leads to 18 occupied and 364 virtual spinors in the correlation space. Table 5 presents the excitation energy and transition moment in 4c-ADC(3) and 4c-EOM-CCSD methods[62] along with the corresponding experimental estimates[63,64]. It can be seen that the excitation energies in 4c-ADC(3) method are consistently underestimated as compared to the 4c-EOM-CCSD results. The maximum absolute deviation is observed for the state $5p^5({}^2P_{3/2})6s\ {}^2[3/2]_1^o$. Whereas for the $5p^5({}^2P_{3/2})5d\ {}^2[3/2]_1^o$ state, the ADC(3) method shows much better agreement with experimental excitation energy that observed for the 4c-EOM-CCSD method. . However, the quality of the calculated transition dipole moments for all the states in 4c-ADC(3) is comparable to their 4c-EOM-CCSD analogue.

The present implementation of 4c-ADC(3) method is equally applicable to atoms and molecules. To investigate the performance 4c-ADC(3) method for excited state properties in molecules, we have chosen the example of hydrogen iodide (HI) molecule. For many years, the HI has generated substantial experimental[65–69] and theoretical interest[70–75]. The ultraviolet photolysis of hydrogen iodide is significant in both photochemical and spectroscopic studies. The HI molecule's unique and complicated dissociation channel made it an interesting system to study. After absorption, the HI molecule immediately dissociates into two dissociation channels, that leads to the generation of two states: the ground $^2P_{3/2}$ (I) and the first excited $^2P_{1/2}$ (I*) state. From a spectroscopic perspective, hydrogen iodide is still quite a mystery. A theoretical investigation by Mulliken[70] assigned the broad peak of UV absorption spectra of HI near 5.64 eV (45,500 cm$^{-1}$) as a result of three transitions, one parallel $^1\Sigma_{0^+} \rightarrow {}^3\Pi_{0^+}$ and two perpendicular $^1\Sigma_{0^+} \rightarrow {}^3\Pi_1$, $^1\Sigma_{0^+} \rightarrow {}^1\Pi_1$ transitions.. The motivation for using relativistic HI molecule as a test case stems from the significant spin-orbit interactions due to the presence of iodine atom. Several investigations are reported in the literature: spin-orbit coupled configuration-interaction (CI) calculations by Buenker and colleagues[71] , an empirical model-based analysis by Brown and collaborators[75], configuration-interaction (CI) calculations utilising relativistic effective core potentials by Chapman et al[72–74] and recently 4c-EOM-CCSD study by our group[62]. Here, we have investigated the first three bright states of HI using 4c-ADC(3) method. From Table 6, One can see that the excitation energies in 4c-ADC(3) method is underestimated as compared to the 4c-EOM-CCSD method, similar to that observed in earlier examples. The 4c-EOM-CCSD and MRD-CI[71] results are in good agreement with each other. These two methods show better agreement with the experiment as compared to the 4c-ADC(3) method. However, comparison of theoretical results calculated with double zeta level with the experiment may not be justified. For the transition moment, 4c-ADC(3) results are in close agreement with the corresponding 4c-EOM-CCSD values. The transition moments of MRD-CI method also show very similar values.

The dipole moments of the HI molecule are critical for understanding its electronic structure and spectroscopic behaviour. In particular, the dipole moments of the ground state, $X\,^1\Sigma^+$, and the excited triplet state, $a\,^3\Pi$, determine the intensity of transitions between these states. These transitions are normally spin-forbidden, but mixing between the $X\,^1\Sigma^+$ ground state and $a\,^3\Pi$ states via spin-orbit coupling allows for these transitions to occur, making them observable in the absorption spectrum, as one can see from the transition moments reported in Table 6. The strength of the transition moment, which dictates the spectral intensity, is proportional to the difference between the dipole moments of the two states. The direct mixing of the $X\,^1\Sigma^+$ and $a\,^3\Pi$ states is the primary mechanism responsible for giving intensity to the $^1\Sigma_{0^+} \rightarrow {}^3\Pi_{0^+}$ transition and the intensity is proportional to the difference in the dipole moments of these states. The calculated absolute values of the dipole moment difference ( $|\boldsymbol{\Delta\mu}| = \boldsymbol{\mu_e} - \boldsymbol{\mu_g}$ ) of the three excited states have been reported in Table 6. At the MRD-CI level, dipole moments are only given for $a\,^3\Pi$ and $A\,^1\Pi$ states without considering spin-orbit coupling. It is evident that the excitation of an electron from the iodine atom’s lone-pair $\pi$ orbital to the anti-bonding $\sigma^*$ orbital, primarily located on the hydrogen atom, causes a major shift in the molecule’s electronic density, leading to a significant change in dipole moment in the excited states. From Table 6, it can be seen that the difference in dipole moment is prominent for all the three excited states with respect to the ground state. Among the three excited states, the $a\,^3\Pi_{0+}$ state shows the highest dipole moment difference in the 4c-ADC(3) method, producing a large contribution to the transition dipole moment.

## 5. Conclusions

We have reported the theory and implementation of a third-order algebraic diagrammatic construction method based on a four-component relativistic Dirac-Coulomb Hamiltonian. The performance of the method has been investigated by calculating ionization potential, electron affinity, excitation energy, and excited state properties of heavy atoms and molecules. The perturbative calculation of the ground state wave function in the 4c-ADC(3) method results in a lower computational cost than the corresponding 4c-EOM-CCSD method. The results obtained using the 4c-ADC(3) method show good agreement with the available experimental values. For IP, EA, EE and excited state properties 4c-ADC(3) shows qualitatively similar results to that of the corresponding 4c-EOM-CCSD method. The main advantage of the 4c-ADC(3) method lies in the property calculation, where due to the Hermitian nature of the similarity transformed Hamiltonian, one needs to calculate only one set of eigenvectors, unlike the EOM-CCSD method, where one needs to calculate the left eigenvectors for the ground and excited state in addition to the right eigenvectors. Therefore, the computational cost of property calculations in the 4c-ADC(3) method is nearly half of the EOM-CCSD method. The ADC(3) method shows a balance between computational cost and accuracy, which is suitable for practical applications. Further cost reduction and more efficient implementation to perform calculations on larger systems are necessary, especially in a relativistic framework, and work is in progress toward that direction.

### Supplementary Material

Programmable expressions for 4c-IP/EA/EE-ADC(3) method have been provided in the Supplementary Material.

## ACKNOWLEDGMENTS

The authors acknowledge the support from IIT Bombay, IIT Bombay Seed Grant (Project No. R.D./0517-IRCCSH0-040), CRG (Project No. CRG/2022/005672) and MATRICS (Project No. MTR/2021/000420) project of DST-SERB, CSIR-India (Project No. 01(3035)/21/EMR-II), DST-Inspire Faculty Fellowship (Project No. DST/INSPIRE/04/2017/001730), Prime Minister's Research Fellowship (PMRF) and ISRO for financial support, IIT Bombay super computational facility, and C-DAC Supercomputing resources (Param Smriti and Param Bramha) for computational time.

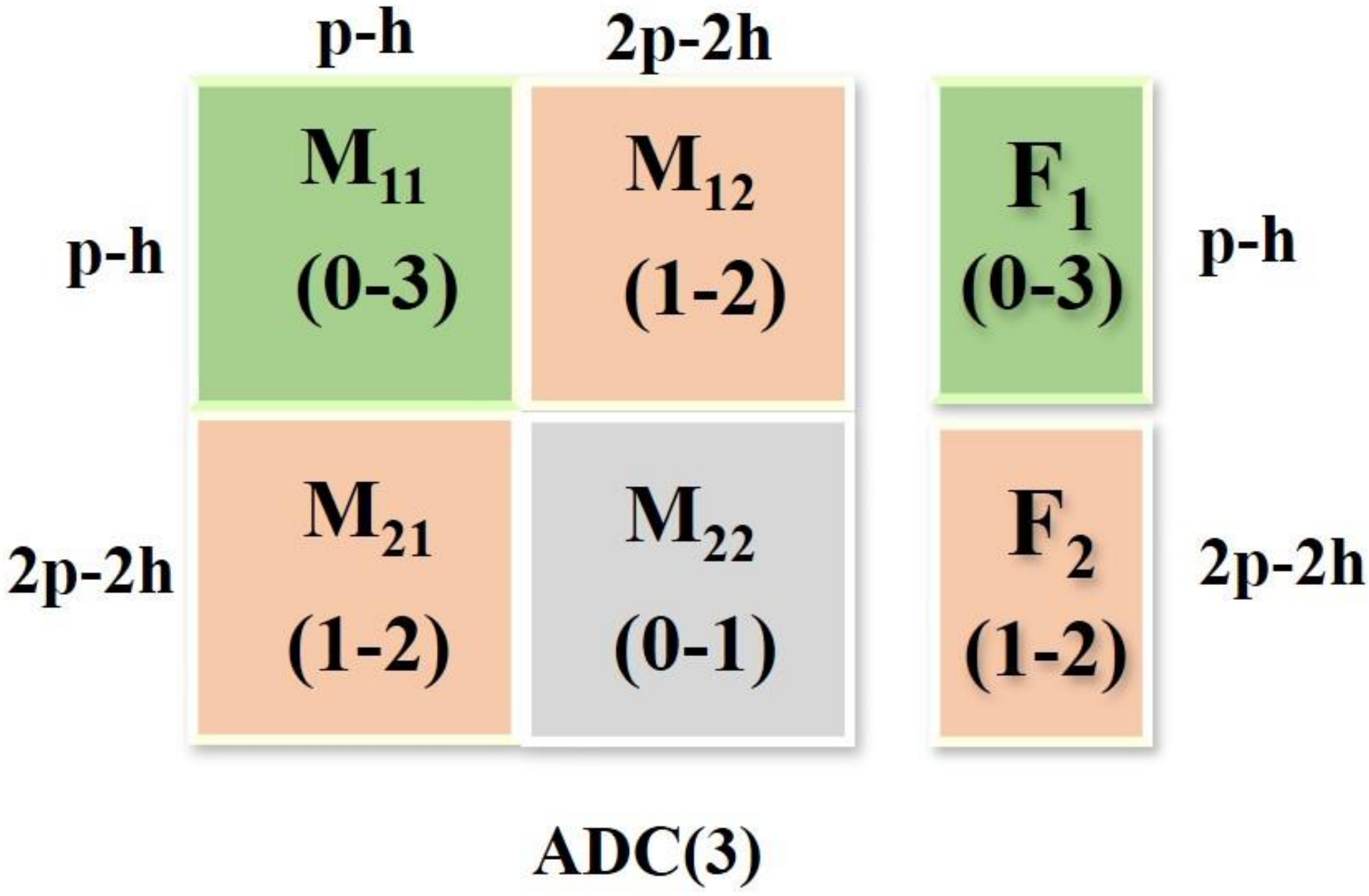

***Figure 1: Block wise of third-order ADC(3) ISR matrix (M) and Trnaision amplitude (F)***

***Table 1: The comparison of splittings of states in the IP values (in eV) calculated with ADC(2) and ADC(3) methods with respect to the experimental results (in eV) for halogen monoxide anions ($XO^-$; X=Cl, Br, I) in the basis Dyall.av3z for X atom and uncontracted aug-cc-pVTZ for O atom.***

| Molecule | Method | $^2\Pi_{3/2}$ | $^2\Pi_{1/2}$ | Splitting | Expt[76]. | Error |
|---|---|---|---|---|---|---|
| $ClO^-$ | 4c-IP-ADC(2) | 0.7779 | 0.8091 | 0.0312 | 0.0397 | -0.0085 |
| | 4c-IP-ADC(3) | 2.4158 | 2.4660 | 0.0503 | | 0.0106 |
| | 4c-IP-ADC[(2)+x(3)] | 1.7446 | 1.7862 | 0.0415 | | 0.0018 |
| | 4c-IP-EOM-CCSD | | | 0.0422 | | 0.0025 |
| $BrO^-$ | 4c-IP-ADC(2) | 0.8701 | 0.9462 | 0.0761 | 0.1270 | -0.0509 |
| | 4c-IP-ADC(3) | 2.5819 | 2.7470 | 0.1651 | | 0.0381 |
| | 4c-IP-ADC[(2)+x(3)] | 1.9125 | 2.0347 | 0.1222 | | -0.0048 |
| | 4c-IP-EOM-CCSD | | | 0.1294 | | 0.0024 |
| $IO^-$ | 4c-IP-ADC(2) | 0.9397 | 1.0697 | 0.1300 | 0.2593 | -0.1293 |
| | 4c-IP-ADC(3) | 2.5889 | 2.9760 | 0.3874 | | 0.1281 |
| | 4c-IP-ADC[(2)+x(3)] | 2.0161 | 2.2791 | 0.2629 | | 0.0036 |
| | 4c-IP-EOM-CCSD | | | 0.2776 | | 0.0183 |

***Table 2: The vertical electron attachment energies (in eV) of halogen atoms (F, Cl, Br, I, and At) obtained using the EA-ADC(3) method and the comparison with the corresponding experimental values (eV).****

| | NR-ADC(3)[a] | SFX2C1e-ADC(3)[a] | 4c-ADC(2) | 4c-ADC(3) | 4c-EOM-CCSD | Expt. |
|---|---|---|---|---|---|---|
| F | 2.886 | 2.875 | 3.103 | 3.018 | 3.227 | 3.401[77] |
| Cl | 3.447 | 3.432 | 3.616 | 3.674 | 3.490 | 3.613[78] |
| Br | 3.391 | 3.347 | 3.337 | 3.346 | 3.221 | 3.363[79] |
| I | 3.289 | 3.214 | 3.067 | 3.116 | 2.929 | 3.059[80] |
| At | 3.102 | 2.920 | 2.495 | 2.362 | 2.283 | 2.416[81] |

*All the calculations are performed in d-aug-Dyall.v4z basis set

*Table 3: Vertical excitation energy (eV) of Triiodide anion ( $I_3^-$ ) using 4c-EE-ADC(3) with Dyall.av3z basis set. The $DC^M$-EOM/IHFS are taken from the work of Shee et al*[54].

| State | Ω | $DC^M$ | | $DC^M$ | | EOM[61] | | | CASPT2[51] |
|---|---|---|---|---|---|---|---|---|---|
| | | ADC(2) | ADC(3) | EOM[54] | IHFS[54] | (a) | (b) | (c) | |
| 1 | $2_g$ | 2.25 | 2.06 | 2.24 | 2.07 | 2.22 | 2.36 | 2.16 | 2.24 |
| 2 | $1_g$ | 2.38 | 2.19 | 2.37 | 2.20 | 2.35 | 2.50 | 2.29 | 2.32 |
| 3 | $0_u^-$ | 2.49 | 2.23 | 2.37 | 2.22 | 2.34 | 2.49 | 2.29 | 2.47 |
| 4 | $1_u$ | 2.50 | 2.23 | 2.38 | 2.23 | 2.34 | 2.47 | 2.30 | 2.47 |
| 5 | $0_g^-$ | 2.85 | 2.66 | 2.84 | 2.66 | 2.81 | 2.96 | 2.72 | 2.76 |
| 6 | $0_g^+$ | 2.90 | 2.71 | 2.89 | 2.71 | 2.86 | 2.99 | 2.75 | 2.82 |
| 7 | $1_g$ | 3.05 | 2.87 | 3.07 | 2.88 | 3.04 | 3.20 | 2.96 | 2.85 |
| 8 | $2_u$ | 3.25 | 3.13 | 3.32 | 3.19 | 3.30 | 3.47 | 3.25 | 3.10 |
| 9 | $1_u$ | 3.32 | 3.19 | 3.41 | 3.27 | 3.39 | 3.55 | 3.34 | 3.11 |
| 10 | $0_u^+$ | 3.57 | 3.43 | 3.66 | 3.51 | 3.65 | 3.79 | 3.56 | 3.52 |
| 11 | $2_g$ | 4.10 | 3.88 | 4.09 | 3.92 | 4.04 | 4.19 | 3.98 | 3.98 |
| 12 | $0_u^-$ | 4.01 | 3.89 | 4.08 | 3.93 | 4.05 | 4.18 | 3.91 | 3.79 |
| 13 | $1_u$ | 4.10 | 3.96 | 4.18 | 4.02 | 4.15 | 4.29 | 4.01 | 3.80 |
| 14 | $1_g$ | 4.23 | 3.99 | 4.21 | 4.03 | 4.17 | 4.32 | 4.10 | 4.06 |
| 15 | $0_u^+$ | 4.40 | 4.28 | 4.49 | 4.33 | 4.50 | 4.67 | 4.42 | 4.51 |
| 16 | $0_g^-$ | 4.70 | 4.49 | 4.69 | 4.51 | 4.65 | 4.76 | 4.51 | 4.51 |
| 17 | $0_g^+$ | 4.72 | 4.50 | 4.70 | 4.51 | 4.65 | 4.82 | 4.51 | 4.53 |
| 18 | $1_g$ | 4.87 | 4.85 | 4.90 | 4.71 | 4.86 | 4.99 | 4.73 | 4.60 |
| MAD | | 0.16 | 0.04 | 0.17 | | 0.13 | 0.28 | 0.05 | 0.11 |
| STD | | 0.07 | 0.04 | 0.02 | | 0.02 | 0.02 | 0.03 | 0.08 |
| RMSD | | 0.17 | 0.05 | 0.17 | | 0.14 | 0.28 | 0.06 | 0.14 |

**(a) SOC operator included in the post Hartree-Fock calculations (b) SOC operator included only in the EOM step (c) The size intensive approximate variant of (b)**

***Table 4: Excitation energies (in eV) and fine-structure splitting (in eV) of $Ga^+$, $In^+$ and $Tl^+$ computed using 4c-ADC(3)/ t-aug-Dyall.v3z level of theory, compared with experiments.***

| | $Ga^+$ | $In^+$ | $Tl^+$ |
|---|---|---|---|
| 4c-EE-ADC(3) | | | |
| $^1S_0 \rightarrow {}^3P_0$ | 5.534 | 4.851 | 5.983 |
| $^1S_0 \rightarrow {}^3P_1$ | 5.589 | 4.982 | 6.344 |
| $^1S_0 \rightarrow {}^3P_2$ | 5.702 | 5.275 | 7.451 |
| $^1S_0 \rightarrow {}^1P_1$ | 8.610 | 7.673 | 9.221 |
| Splitting | | | |
| $^3P_0 \rightarrow {}^3P_1$ | 0.054 | 0.131 | 0.361 |
| $^3P_1 \rightarrow {}^3P_2$ | 0.113 | 0.293 | 1.107 |
| Experiment | | | |
| $^1S_0 \rightarrow {}^3P_0$ | 5.874 | 5.242 | 6.131 |
| $^1S_0 \rightarrow {}^3P_1$ | 5.928 | 5.375 | 6.496 |
| $^1S_0 \rightarrow {}^3P_2$ | 6.044 | 5.682 | 7.653 |
| $^1S_0 \rightarrow {}^1P_1$ | 8.766 | 7.816 | 9.381 |
| Splitting | | | |
| $^3P_0 \rightarrow {}^3P_1$ | 0.055 | 0.133 | 0.365 |
| $^3P_1 \rightarrow {}^3P_2$ | 0.116 | 0.307 | 1.157 |

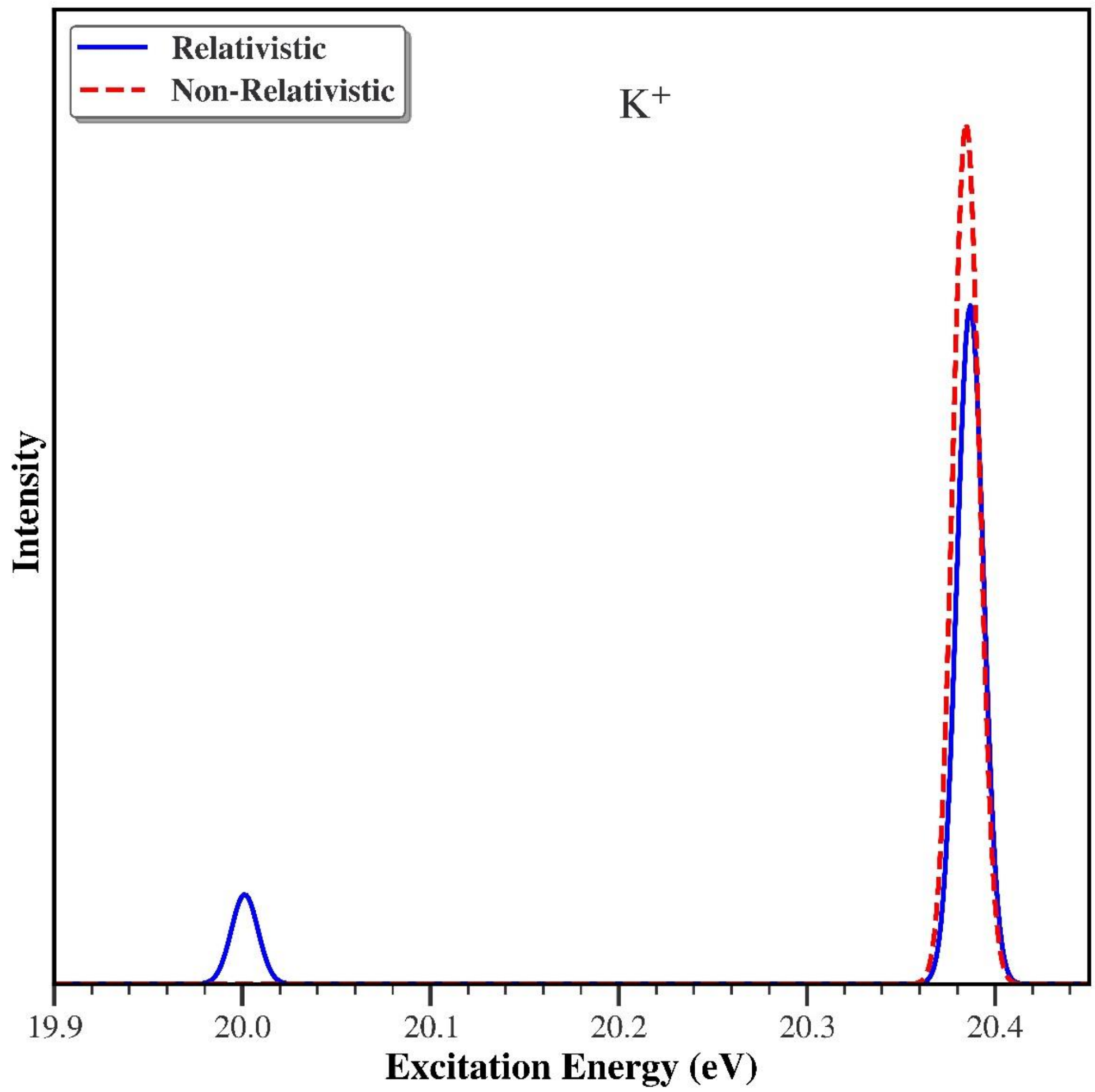

***Figure 2: Absorption spectra of $K^+$ ion computed using 4c-EE-ADC(3) and non-relativistic EE-ADC(3) using s-aug-Dyall.v2z basis set.***

***Table 5: Excitation energy (in eV) and Transition Dipole Moments (TDM) (in a.u.) of Xe atom for the first four excited states in ADC(3) and EOM-CCSD method. d-aug-Dyall.ae3z basis set is used.***

| | 4c-EOM-CCSD[a] | | 4c-ADC(3) | | Experiment[63] | |
|---|---|---|---|---|---|---|
| Excited state | EE | TDM | EE | TDM | EE | TDM |
| $5p^5(^2P_{3/2})6s\ ^2[3/2]^o_1$ | 8.414 | 0.647 | 8.172 | 0.643 | 8.436 | 0.654 |
| $5p^5(^2P_{1/2})6s\ ^2[1/2]^o_1$ | 9.567 | 0.535 | 9.308 | 0.522 | 9.569 | 0.521 |
| $5p^5(^2P_{3/2})5d\ ^2[1/2]^o_1$ | 9.964 | 0.058 | 9.697 | 0.047 | 9.917 | 0.120 |
| $5p^5(^2P_{3/2})5d\ ^2[3/2]^o_1$ | 10.628 | 0.949 | 10.417 | 0.892 | 10.401 | 0.704 |

[a] taken from ref[62]

***Table 6: Excitation Energy (EE in eV) and Transition Dipole Moment (TDM in a.u.) and change in dipole moment ($|\Delta\mu| = \mu_e - \mu_g$ in a.u.) for different excitations from the ground state ($X\ ^1\Sigma^+$) of HI molecule.***

| State | 4c-ADC(3) | | | 4c-EOM-CCSD | | MRD-CI | | | Exp |
|---|---|---|---|---|---|---|---|---|---|
| | EE[a] | TDM[a] | $\Delta\mu$ | EE[a] | TDM[a] | EE | TDM | $\Delta\mu$ | EE |
| $a\ ^3\Pi_1$ | 4.76 | 0.0692 | 0.4792 | 5.02 | 0.0626 | 4.98 | 0.0654 | 0.5724 | 5.04 |
| $a\ ^3\Pi_{0+}$ | 5.25 | 0.1617 | 0.4884 | 5.50 | 0.1631 | 5.45 | 0.1519 | - | 5.47 |
| $A\ ^1\Pi_1$ | 5.50 | 0.1248 | 0.4427 | 5.77 | 0.1343 | 5.74 | 0.1538 | 0.5613 | 5.96 |

[a] Calculated using d-aug-Dyall.ae2z